\documentclass[preprint]{aastex6}
\usepackage{natbib}
\begin{document}

\title{A Correction for GALEX Spectroscopic UV Flux Distributions from
Comparisons with  CALSPEC and IUE}

\author{Ralph~C.\ Bohlin\altaffilmark{1}, and Luciana Bianchi\altaffilmark{2}}
\altaffiltext{1}{Space Telescope Science Institute, 3700 San Martin Drive,
Baltimore,  MD 21218, USA}
\altaffiltext{2}{Department of Physics and Astronomy, The Johns Hopkins
University, Baltimore, MD 21218, USA}

\begin{abstract}

The CALSPEC database of absolute spectral energy distributions (SEDs)
from the \emph{Hubble Space Telescope} (HST) is based on pure hydrogen model
atmosphere calculations for three unreddened white dwarf (WD) stars and 
represents the current UV flux calibration standard with a precision approaching
1\% for well observed stars. Following our previous work to correct IUE fluxes,
this paper provides an average correction for the GALEX spectral database that
places GALEX fluxes on the current CALSPEC scale. Our correction is derived by
comparing GALEX spectroscopic flux distributions with CALSPEC and
corrected IUE SEDs. This re-calibration is relevant for any project based on
GALEX archival spectroscopic data, e.g. UV or multi-wavelength analyses,
correlating GALEX spectra with other existing or future databases, and planning
of new observations. The re-calibration will be applied to our planned catalog
of corrected GALEX SEDs. \end{abstract}

\keywords{stars: atmospheres --- stars: fundamental parameters
--- techniques: spectroscopic --- ultraviolet: stars}

\section{Introduction}			
\label{s_intro}

Accurate absolute stellar fluxes as a function of wavelength are required for
many astrophysical purposes. The fundamental parameters of stars, including
mass, radius, metallicity, and age are inferred by matching accurate stellar
atmosphere models to precisely calibrated UV spectroscopic data from which the
effective temperature, surface gravity, composition, and interstellar reddening
are determined for all types of hot stellar objects, e.g.
\citet{bianchigarcia02, bianchigarcia14, bianchi2012pn, bianchi2018b} and
references therein: \citet{heraldbianchi07, heraldbianchi11, garciabianchi04,
pala2015, joyce2018}. The set of $>$100,000 GALEX UV spectra with a homogeneous
spectral range of $\sim$1300-3000\AA) and resolution of 8\AA\ in the
far-ultraviolet (FUV) and 20\AA\ in the near-ultraviolet (NUV) is a resource
with similar characterstics to the IUE spectral database but with about ten
times larger sample and fainter fluxes. UV extinction curves can be derived for
up to 1000 sight-lines within the Milky Way, yielding unique information on
properties of interstellar dust and producing extinction maps from a larger
sample (Bianchi, L. et al. in preparation) than previous studies from the IUE
spectral collection, e.g. \citet{valencic2004} with about 400 extinction curves.

\citet{bertone2011} used an extensive sample of early GALEX stellar
spectra to measure stellar line indices for classification. Their work
demonstrated the potential of the GALEX spectral database in spite of
limitations of the flux calibration and of the initial spectral extraction
software, which has been subsequently improved. \citet{bianchi2012pn2} showed
the potential of grism spectra for studying extended objects with UV emission
lines. GALEX grism spectra have been used to study different types of stellar
objects, e.g. \citet{montez2017,godon2014, Gal-Yam2008}, and of extra-galactic
sources \citep{Burgarella2005}, including Ly$\alpha$ emitters
\citep{Deharveng2008, Cowie2010, Wold2017, Barger2012} . The latter two works
extracted a spectral sample of Ly$\alpha$ emitters that is fainter than the
GALEX pipeline extraction limits by stacking the grism images taken at different
orientations, as in the standard pipeline procedure, but searching for line
emission, regardless of a detectable continuum. These examples illustrate the
broad range of topics that can make use of the GALEX serendipitous spectral
collection and underscore the importance of a calibration consistent with the
current standard flux scale.

Critical for any study using GALEX spectra  is the verification of the
flux calibration of GALEX spectra from the cross-strapping of hot white dwarfs
of intermediate brightness between GALEX and HST/IUE UV calibration standards.
The resulting recalibration that is presented in this work will support any
investigation using GALEX spectra and will allow more meaningful comparisons
with models or with observations of the same targets from other instruments.
Therefore, this updated calibration is available here, before our release of
a major spectral database with source classification \citep{bianchi2018}.

Currently, the best choice of fundamental standards in the UV to near-IR seems
to be the CALSPEC models for the primary pure hydrogen white dwarfs (WDs)
G191B2B, GD153, and GD71. Our previous publication \citep{b&b2018} specifies
the correction required to place IUE spectral energy distributions (SEDs) on
this
HST/CALSPEC\footnote{http://www.stsci.edu/hst/observatory/crds/calspec.html}
absolute flux scale of \citet{bohlinetal14}. 

\section{GALEX Spectra}		
\label{s_galex}

The GALEX spacecraft was launched on 2003 April 28 and had 10 years of success
in obtaining broadband UV photometry at effective wavelengths of 1539 (FUV) and
2316~\AA\ (NUV), in addition to spectrophotometry with a CaF$_2$ grism that is
sampled at 3.5~\AA\ intervals from 1300 to 3000~\AA. The spectral resolution is
8~\AA\ in the far-ultraviolet (FUV) second order and 20~\AA\ in the
near-ultraviolet (NUV) first order with a gap near 1820~\AA\ between the FUV and
NUV ranges \citep{morrissey2007}. Each grism observation was obtained with
several different orientations; and for each source identified in the
corresponding direct image in each detector (FUV and NUV), the portions of the
source spectrum not contaminated by overlapping spectra of nearby sources in
each orientation sub-exposure are extracted (see figure 1 of
\citet{bianchi2018}). All the non-contaminated segments are coadded by the
pipeline into a 2-dimensional spectral ``strip'' from which the spectral flux
and background are extracted and calibrated, resulting in a one-dimensional
(wavelength, flux, sigma) spectrum for each source in an observation. These
one-dimensional GALEX final extracted spectral \textit{*.fits} files are the
only GALEX data products analyzed in this paper. The exposure time is not
necessarily uniform even across one single spectrum, because each archival SED
is made up of all segments that are not contaminated by overlapping spectra of
nearby sources in the separate exposures at the several orientations that
comprise one observation \citep{morrissey2007}. The effect of varying exposure
time within each spectrum is captured by the statistical error array (sigma)
that accompanies the GALEX final extracted flux distribution. Our analysis
always weights each GALEX sample point by its statistical uncertainty.

Both the CALSPEC and IUE\footnote{http://archive.stsci.edu/iue/} data sets are
compared with GALEX\footnote{http://galex.stsci.edu/GR6/} spectral flux
distributions in preparation for a re-calibration and publication of the
relatively unexploited 125,564 GALEX prism spectra, which comprise a homogeneous
set fainter and tenfold larger than the IUE sample (\citet{bianchi2018}, Bianchi
et al. in prep). Our new database will facilitate future use of GALEX
spectra for a variety of purposes.

Each GALEX match is verified during both sets of comparisons, i.e. with the few
overlapping CALSPEC stars in Section \ref{s_calspec} and with several IUE SEDs
in Section \ref{s_iue}. Section \ref{s_result} combines the results of the
previous two sections to derive a final best average update to the
GALEX flux calibration, while Section \ref{s_el} details the electronic access
to our results.

\section{Comparison of GALEX and CALSPEC} 
\label{s_calspec}

\begin{figure}			
\centering 
\includegraphics*[height=6.5in]{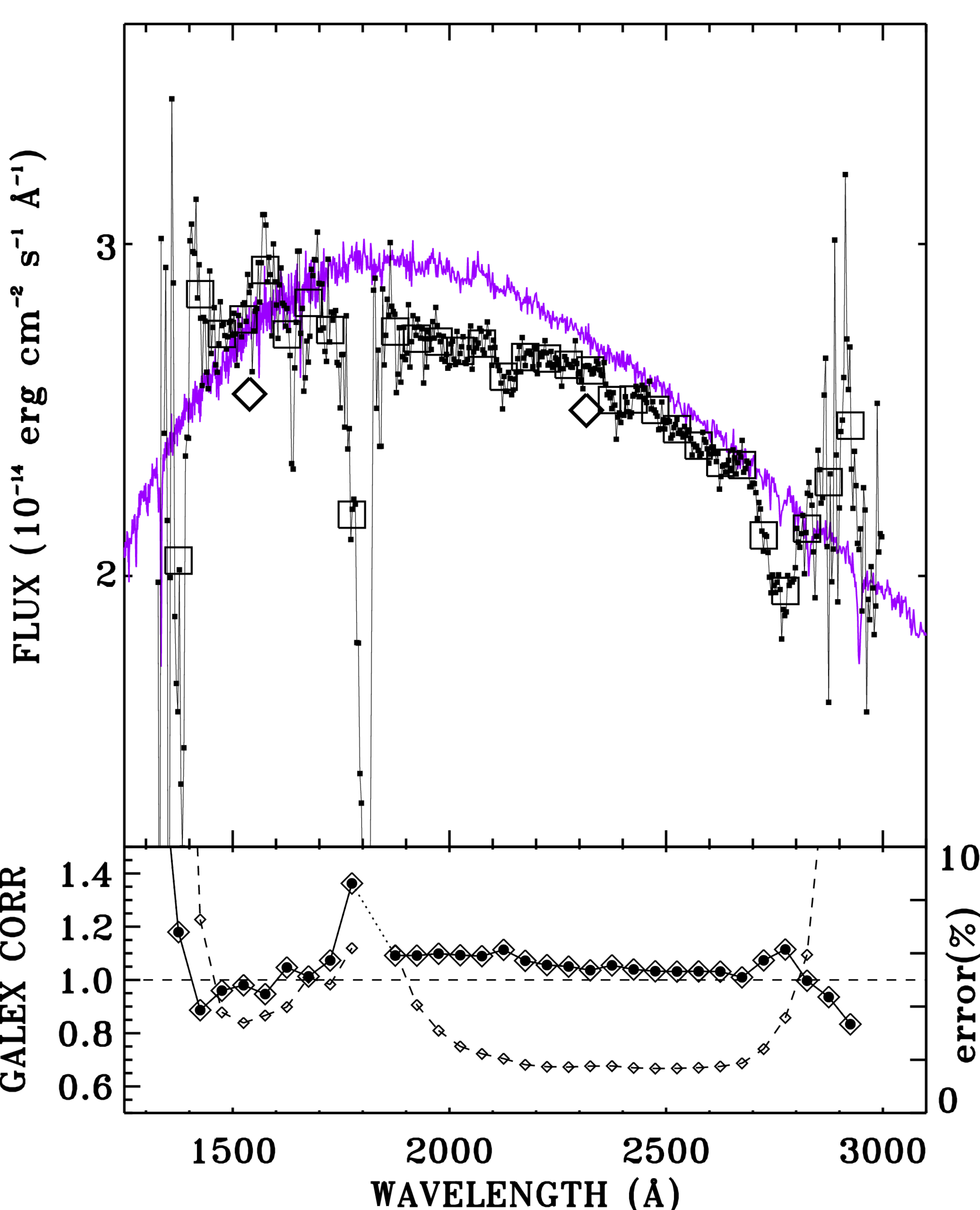}
\caption{\baselineskip=12pt
CALSPEC spectrum of LDS749B (purple) overplotted with the GALEX spectrum as
small black squares and as large black squares in 50~\AA\ bins. The large black
diamonds are the $\lambda_{eff}$=1539 and 2316~\AA\ GALEX broad-band photometry 
that is within 10\% of the spectral measures. In the lower panel, the ratio of
CALSPEC/GALEX is shown as black circles surrounded by large black diamonds. The
GALEX plus CALSPEC statistical uncertainties combined in quadrature appear as
small black diamonds connected by a dashed line with a scale in percent
indicated on the right axis. \label{clsplds}} \end{figure}

\begin{figure}			
\centering 
\includegraphics*[height=6.5in]{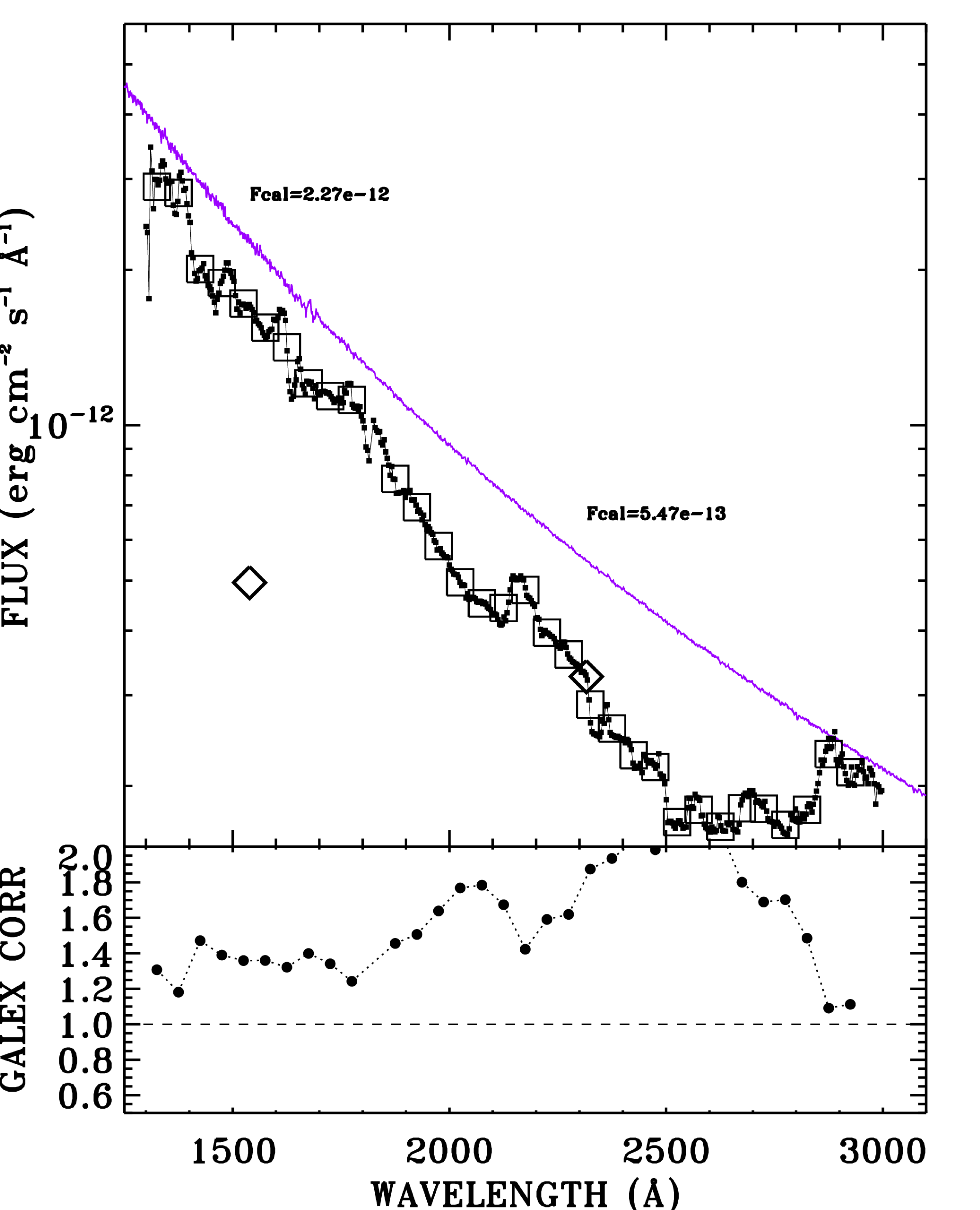}
\caption{\baselineskip=12pt
CALSPEC spectrum of HZ43 (purple) overplotted with the GALEX spectrum as small
black squares and as large black squares in 50~\AA\ bins, as in
Figure~\ref{clsplds}. The flux values of the CALSPEC SED at the 1539 and
2316~\AA\ reference wavelengths are written on the plot. Both the GALEX
photometry and spectral fluxes are low, as expected when the  GALEX detector
approaches saturation. In the lower panel, there are no little or big black
diamonds that indicate useful measures of the CALSPEC/GALEX flux ratio.
\label{clsphz43}} \end{figure}

GALEX spectra exist for 11 CALSPEC stars, and Table~\ref{table:calsp}
summarizes the 11 matches. Figure~\ref{clsplds} illustrates the comparison of
the GALEX and CALSPEC SEDs for the GALEX primary calibration star LDS749B
\citep{bohlin-k2008}. However, the GALEX data of many of the 11 CALSPEC stars
show saturation, as illustrated for HZ43 in Figure~\ref{clsphz43}. In general,
GALEX spectra have some saturation when the true stellar flux is more than
4.5e$^{-13}$ erg cm$^{-2}$ s$^{-1}$ \AA$^{-1}$ in either the FUV or NUV near
the GALEX broadband-photometry effective-wavelengths of 1539 and 2316~\AA.
These two reference wavelengths lie within the regions of the greatest
spectrophotometric sensitivity for the GALEX effective
area\footnote{https://asd.gsfc.nasa.gov/archive/galex/Documents/PostLaunchResponseCurveData.html}.
The effective area is within 10\% of peak from 1470--1550 for the FUV and
2150--2630~\AA\ for the NUV.

The GALEX delay-line type of detector is a photoelectron-counting,
microchannel-plate device that is similar to the COS detector on HST. Due to
inherent limits on how fast the electronics can count pulses, a 10\%
non-linearity (i.e, saturation of the pulse-counting circuits) occurs at a
global rate of 18,000 counts s$^{-1}$ for both the FUV and NUV detectors or at a
total local count rate of 114 and 303 counts s$^{-1}$ for the FUV and NUV,
respectively, for point source photometry \citep{morrissey2007}. According to
the encircled energy curves of \citet{morrissey2007}, the central  pixel
(1.5\arcsec\ x 1.5\arcsec\ size) encloses $\sim$10\% of the total, so the bright
limit for a single pixel is about 11 and 30 counts s$^{-1}$ for FUV and NUV,
respectively. This local count rate limit arises because of the high
$\sim$10$^{7}$ gain of the microchannel
plates\footnote{\url{http://www.galex.caltech.edu/researcher/techdoc-ch1.html\#2}}
and consequent limit on the local rate of charge extraction from the cathodes
without causing damage. For spectrophotometry, the total system peak effective
areas of 20 (FUV) and 40 (NUV) cm$^2$ determine the count rates for our 
4.5e$^{-13}$ erg cm$^{-2}$ s$^{-1}$ \AA$^{-1}$ bright limit for both the FUV and
NUV. Using spectral widths of 2.46 (FUV) and 6.06 (NUV) \AA~pixel$^{-1}$, the
corresponding count~s$^{-1}$ rates for an infinite extraction width
perpendicular to the dispersion are 1.7 (FUV) and 12.7 (NUV), which are
conservative and well below the 11 and 30 count rate limits for a 10\%
non-linearity and single pixel area. For a single spectrum of a hot star with
our brightness limit, the total grism  count rate is also well below the 18,000
global rate limit. For example, the FWHM of the NUV  GALEX effective area is
$\sim$700~\AA, the number of pixels at 6.06 \AA~pixel$^{-1}$ is 116, and the
total count rate is of order 116*12.7, i.e. $\sim$1500~count~s$^{-1}$.

\newpage
\begin{deluxetable}{llccl}    
\tablewidth{0pt}
\tablecolumns{5}
\tablecaption{\label{table:calsp} GALEX Spectra with CALSPEC SEDs}
\tablehead{
\colhead{Star} &\colhead{CALSPEC Name} &\colhead{FUV\tablenotemark{a}} 
	&\colhead{NUV\tablenotemark{a}} &\colhead{Comment}}
\startdata
 BD+33$^{\circ}$2642  &bd\_33d2642\_fos\_003     &28.0 &13.5  &Saturation	\\
    G191B2B  &g191b2b\_mod\_010       &70.1 &16.5  &Saturation	\\
      GD153  &gd153\_mod\_010 	      &13.3 &3.34  &FUV Saturation	\\
GRW+70$^{\circ}$5824  &grw\_70d5824\_stisnic\_007 &8.81 &2.75  &FUV Saturation, NUV Outlier	\\
        HZ4  &hz4\_stis\_005.fits     &...  &0.31  &No STIS or FOS at $<$1850~\AA\ \\
       HZ21  &hz21\_stis\_004 	      &4.10 &1.08  &Good, 1625~\AA\ bin omitted	 \\
       HZ43  &hz43\_stis\_004 	      &22.7 &5.47  &Saturation	 \\
       HZ44  &hz44\_stis\_004         &43.1 &15.9  &Saturation	 \\
    LDS749B  &lds749b\_stisnic\_006   &0.27 &0.27  &Good	 \\
      P177D  &p177d\_stisnic\_007     &...  &0.004 &No CALSPEC $<$2222~\AA\, Noisy Outlier\\
      P330E  &p330e\_stisnic\_008     &...  &0.009 &No CALSPEC $<$2000~\AA\, Noisy Outlier\\
\enddata
\tablenotetext{a}{10$^{-13} $erg s$^{-1}$ cm$^{-2}$ \AA$^{-1}$ for the
CALSPEC SED at the reference wavelengths of 1539 and 2316~\AA.}
\end{deluxetable}

The exact count rates for saturation depend not only on the source flux but
also on the brightness of other objects in the field-of-view (FOV), on how close
the spectrum is to the detector edge, and possibly on the exact thermal
conditions. While detailed predictions of the amount of saturation are not
within the scope of this paper, our guideline of a 4.5e$^{-13}$ erg cm$^{-2}$
s$^{-1}$ \AA$^{-1}$ non-linearity limit for both the FUV at 1539~\AA\ and the
NUV at 2316~\AA\ is determined by comparing the GALEX spectral flux with the
reference CALSPEC or IUE SED. For example, the IUE 1539~\AA\ spectrophotometric
flux of GD108 is 4.61e$^{-13}$ erg cm$^{-2}$ s$^{-1}$ \AA$^{-1}$, and the GALEX
SED agrees poorly with IUE at some wavelengths.  Conversely, the corresponding
IUE flux for HZ21 is 4.31e$^{-13}$ erg cm$^{-2}$ s$^{-1}$ \AA$^{-1}$, and IUE
agrees with the GALEX flux within 5\% from 1325 to 1775~\AA. The LDS747B flux of
Figure~\ref{clsplds} is below 4.5e$^{-13}$ erg cm$^{-2}$ s$^{-1}$ \AA$^{-1}$ in
both the FUV and NUV, while the HZ43 GALEX SED in Figure~\ref{clsphz43} is
expected to be low due to some saturation. Occasionally for a reference flux
slightly above 4.5e$^{-13}$, a GALEX SED will have minimal saturation losses in
the archival SED; but these cases are excluded for consistency.

The GALEX broadband photometry of HZ43 is even more saturated than the
GALEX spectral data, as can be expected. For the above local 10\% non-linearity
count rate limits, the corresponding effective flux of the broadband photometry
for the zeropoints of \citet{morrissey2007} is about 1.6e$^{-13}$ at 1539~\AA\
and 5.7e$^{-14}$ at 2316~\AA\ in erg s$^{-1}$ cm$^{-2}$ \AA$^{-1}$ units. In
Figure~\ref{clsplds}, the GALEX broadband photometry (diamonds) and the GALEX
spectrophotometry (squares) are both below their respective saturation limits,
and their good agreement confirms that both the photometry and spectrophotometry
have minimal non-linearities.

\begin{figure}			
\centering 
\includegraphics*[height=6.5in]{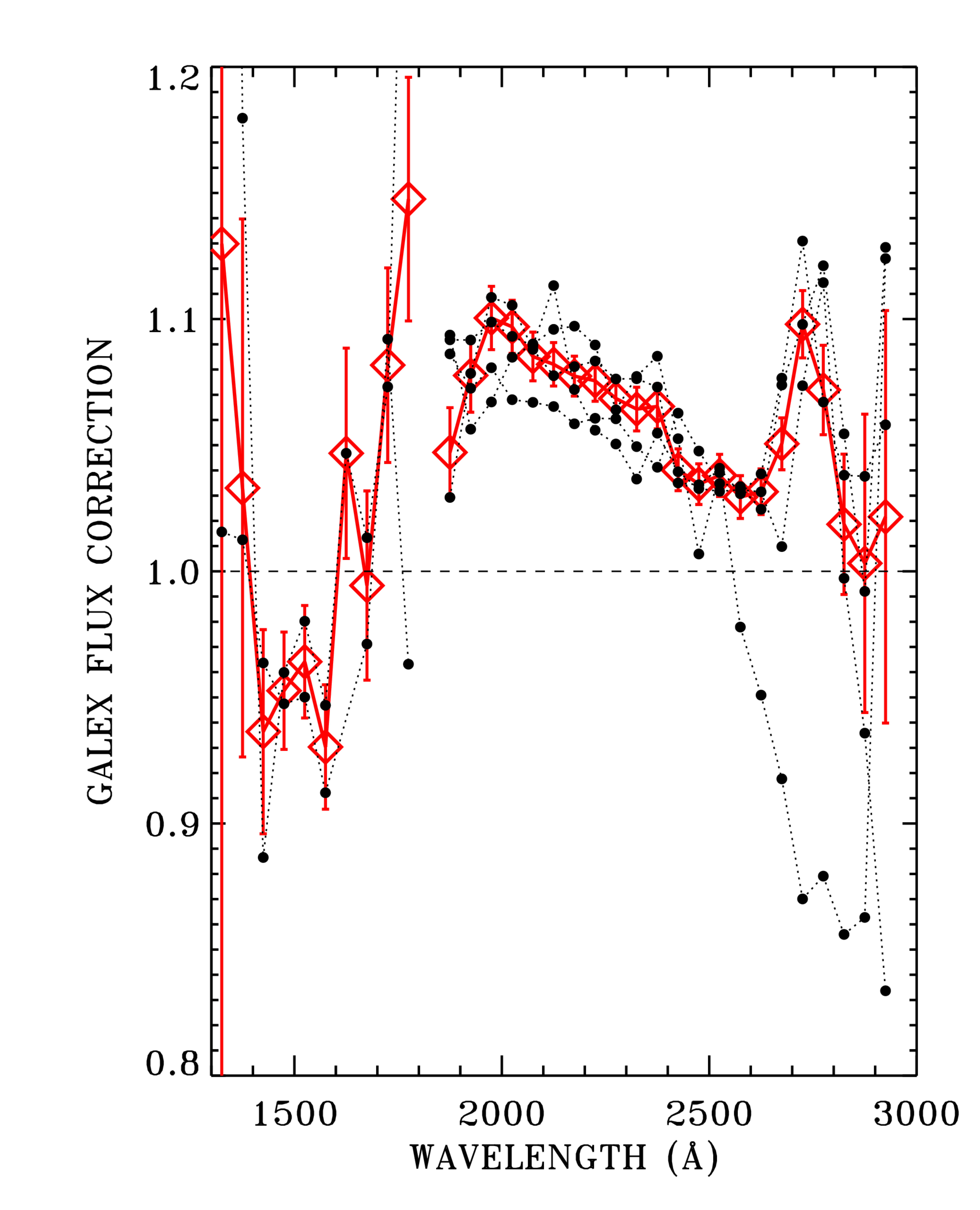}
\caption{\baselineskip=12pt
Average multiplicative correction of GALEX fluxes from direct comparison to
CALSPEC SEDs. Black data points connected by dotted lines are the results from
the two matched stars in the FUV and four stars in the NUV, while the red
diamonds connected by solid lines are the weighted average correction. Results
for HZ4 dive down below unity longward of 2575~\AA\ but have little weight.
Error bars reflect the combination in quadrature of the GALEX and CALSPEC
statistical uncertainties.
\label{clspavg}} \end{figure}

As detailed in Table~\ref{table:calsp}, the GALEX spectra of four stars are
saturated in both channels, and the two G stars P177D and P330E are too faint
and noisy. Two more stars are saturated in the FUV channel, while there are no
CALSPEC data below 1850~\AA\ for HZ4. The GALEX flux for GRW+70$^{\circ}$5824
is almost a factor of two below our adopted saturation limit in the NUV but is
still lower than CALSPEC by more than 10\% and is excluded as an anomaly.
Remaining to define the average CALSPEC/GALEX flux ratio are HZ21 and LDS749B
in the FUV and GD153, HZ4, HZ21, and LDS749B in the NUV, as illustrated in
Figure~\ref{clspavg}, where the red diamonds are the weighted average of the
results for the individual stars. As discussed in \citet{b&b2018}, the
1625~\AA\ bin for HZ21 is omitted because of the variable 1640.5~\AA\ HeII line
that contaminates the ratio for that 50~\AA\ wide bin. The GALEX IDs for the
useful CALSPEC comparisons appear in Table~\ref{table:galx}, along with the
useful IUE data that are discussed in the next Section.

\section{Comparison of GALEX with Corrected IUE Fluxes} 	
\label{s_iue}

\begin{figure}			
\centering 
\includegraphics*[height=6.5in]{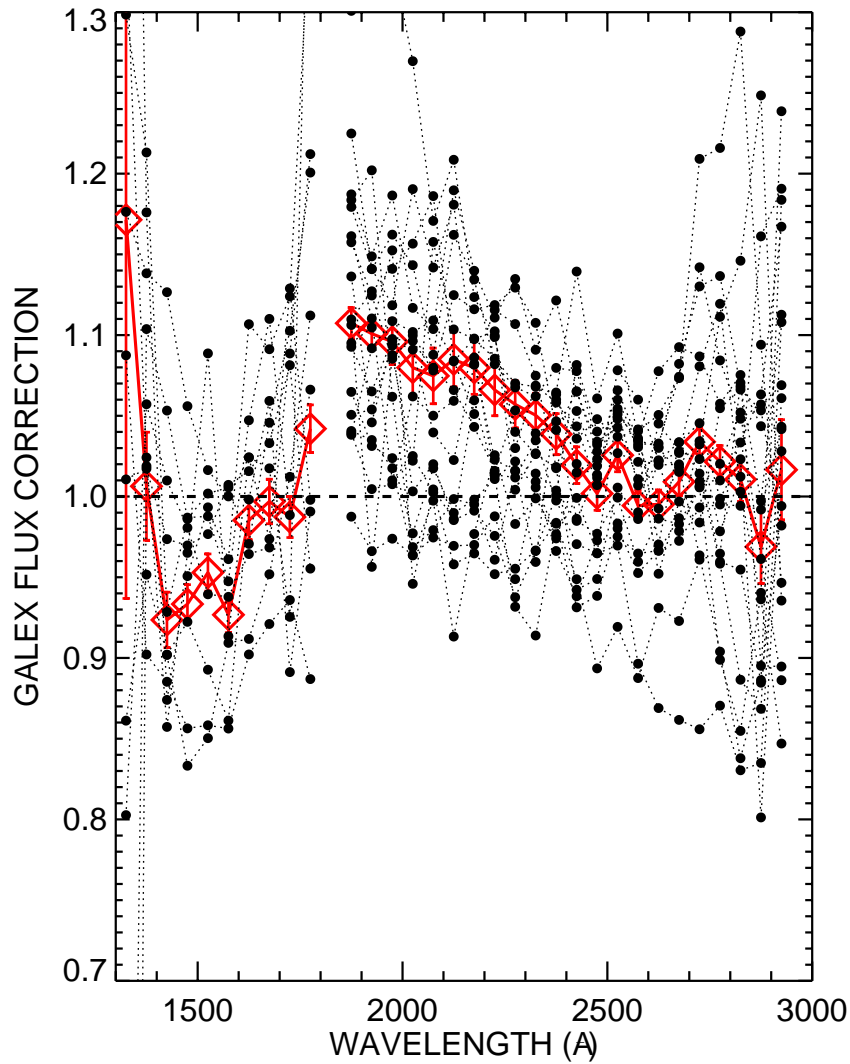}
\caption{\baselineskip=12pt
Average multiplicative correction of GALEX fluxes from comparisons to
IUE SEDs. Black data points connected by dotted lines are the results from
the 10 matched stars in the FUV and 18 stars in the NUV, while the red
diamonds connected by solid lines are the weighted average correction. 
Error bars reflect the combination in quadrature of the GALEX and IUE
statistical uncertainties.
\label{iueavg}} \end{figure}

Even though IUE data are less precise than the CALSPEC data from the Space
Telescope Imaging Spectrograph (STIS) or Faint Object Spectrograph (FOS), there
are many more IUE matches with GALEX spectra than there are CALSPEC matches, so
the GALEX flux correction may improve by using IUE data on the CALSPEC flux
scale \citep{b&b2018}. The IUE vs. GALEX matches are identified by a detailed
cross-match of the two databases that can be found on the $uvsky$ project
website\footnote{\url{http://dolomiti.pha.jhu.edu/uvsky/}} and will be described
further by Bianchi et al. (in prep). In order to find matches of GALEX spectra
with IUE, both databases are downloaded from the MAST archive and are searched
for matches with a relatively large 30\arcsec\ radius, because the  IUE
coordinates are often imprecise. The IUE NEWSIPS *.MXLO files
\citep{nichols1996} for the matched GALEX sources are from the MAST archive and
are co-added and merged as in \citet{b&b2018}. After eliminating spurious
matches, variable stars, and cases where the IUE spectra are mostly noise, there
are 62 potentially useful matches, of which 52 FUV and 44 NUV spectra are
brighter than the GALEX saturation limits, have poor data quality, or have no
IUE match in either the FUV or in the NUV. As collated in
Table~\ref{table:galx}, there remain 10  FUV and 18 NUV matches to define the
average GALEX flux correction, as illustrated in Figure~\ref{iueavg}. All
matches have only one good GALEX observation, except for HZ4 and Q1302-102,
which have two good GALEX SEDs. Table~\ref{table:galx} contains the star names,
coordinates, the IUE stellar flux in the 1539 and 2316~\AA\ regions with the
number of IUE observations in parentheses, and the GALEX spectral data IDs. For
the eight cases lacking FUV measures of the IUE/GALEX flux ratio, the reason for
the lack is indicated as GALEX saturation or poor IUE data quality, preceding
the GALEX ID.

\begin{deluxetable}{lccccl}    
\tablewidth{0pt}
\tablecolumns{6}
\tablecaption{\label{table:galx} Spectra Used for the IUE/GALEX Comparison}
\tablehead{
\colhead{Star} &\colhead{RA} &\colhead{DEC} &\colhead{SWP\tablenotemark{a}} 
	 &\colhead{LW\tablenotemark{a}} &\colhead{GALEX ID}}
\startdata
BPM16274 & 00 50 03.7& -52 08 16& 1.93 (7)& 0.64 (6)&3363524954127600797  \\
SK194	 & 01 45 03.8& -74 31 33& 1.68 (2)& 1.36 (2)&Poor Data,3069735447185853590 \\
SK196	 & 01 49 12.6& -74 00 37& 2.00 (2)& 1.49 (2)&Poor Data,3069735447185857034 \\
GD50	 & 03 48 50.2& -00 58 32& 7.34 (8)& 1.82 (3)&Sat.,3363595322871777529 \\
HZ4	 & 03 55 22.0& +09 47 18& 0.74 (3)& 0.30 (4)&3363665691615955843,3365847122685462904\\
LB227	 & 04 09 28.9& +17 07 54& 0.44 (5)& 0.17 (7)&3363771244732219841 \\
HZ2	 & 04 12 43.6& +11 51 49& 3.29 (5)& 0.98 (5)&3363806429104310591 \\
WD0416-550& 04 17 11.4& -54 57 47& 1.64 (1)& 0.42 (1)&3065126294442221024 \\     
HG7-233  & 04 28 39.4& +16 58 12& 3.56 (1)& 1.08 (1)&3125608230062916337 \\	
GD108	 & 10 00 47.3& -07 33 31& 4.61 (3)& 1.48 (3)&Sat.,3364158272825200648 \\
HZ21	 & 12 13 56.2& +32 56 31& 4.31 (17)&1.13 (13)&3364299010313555707         \\
FEIGE59  & 12 17 21.7& +15 34 58& 2.09 (2)& 1.26 (2)&3061924516582135538 \\
GD153    & 12 57 02.3& +22 01 53& 13.4 (11)&3.30 (10)&Sat.,3364510116546092547 \\
Q1302-102& 13 05 33.0& -10 33 19& 0.33 (7)& 0.12 (4)&3167899845313694736,3186864221869575184\\
LDS749B  & 21 32 16.2& +00 15 14& 0.27 (4)& 0.27 (4)&3364861960266984897  \\
StHA190	 & 21 41 44.9& +02 43 54& 9.90 (3)& 2.75 (1)&Sat.,3192951118238843864 \\	
G93-48   & 21 52 25.4& +02 23 20& 6.12 (3)& 2.20 (3)&Sat.,3365002697755334688 \\
L791-40  & 23 19 35.4& -17 05 28& 0.17 (2)& 0.33 (2)&Poor Data,3365213803987864904 \\
\enddata
\tablenotetext{a}{10$^{-13} $erg s$^{-1}$ cm$^{-2}$ \AA$^{-1}$ and in parentheses are
the number of co-added IUE spectra.}
\end{deluxetable}

\begin{figure}			
\centering 
\includegraphics*[height=6.5in]{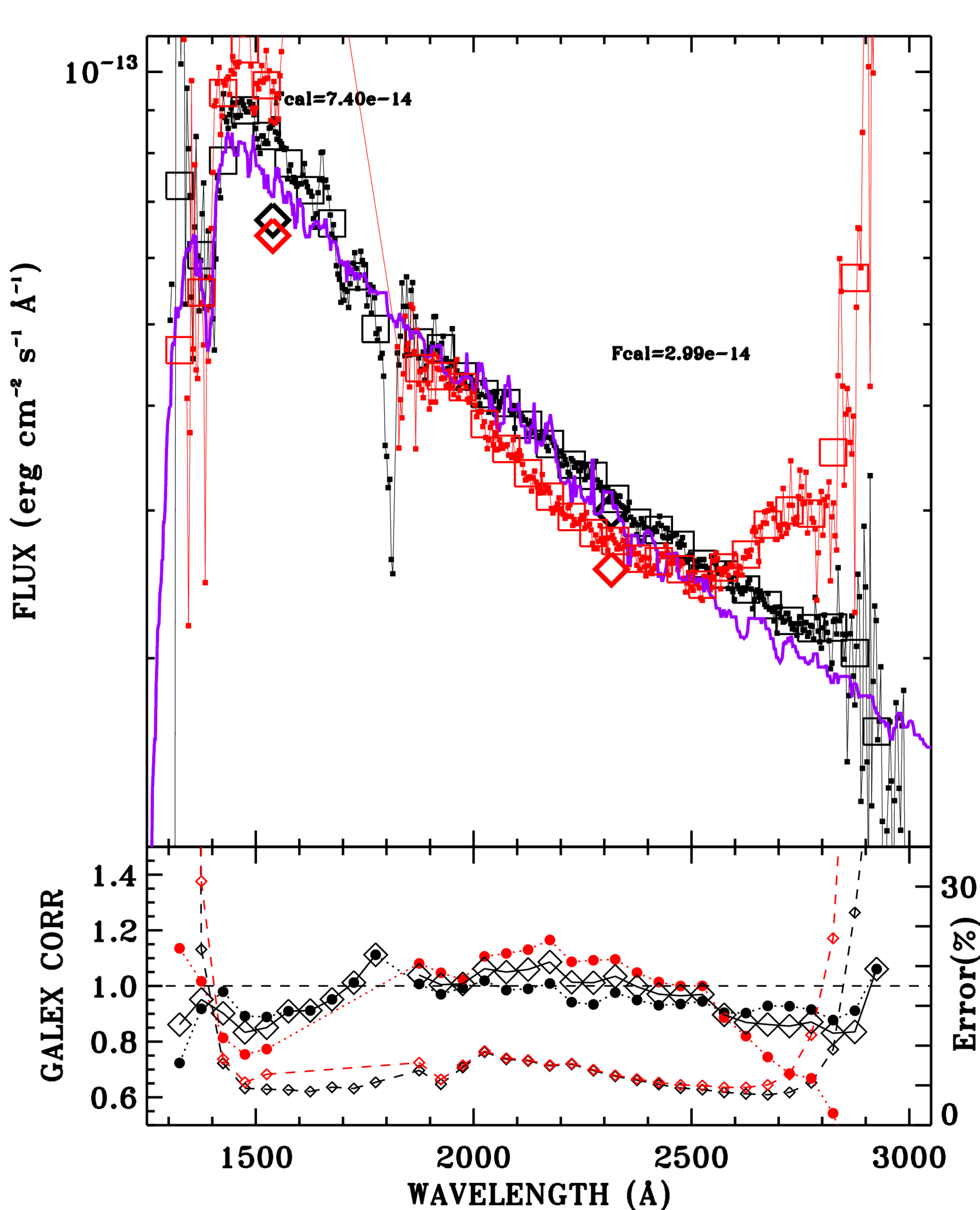} 
\caption{\baselineskip=12pt IUE
spectrum of HZ4 (purple) overplotted with GALEX spectra as small squares and as
large squares in 50~\AA\ bins. The black squares are the GALEX dataset
3363665691615955843, while the red squares are 3365847122685462904. The large
black and red diamonds are the 1539 and 2316~\AA\ GALEX photometry that agrees
fairly well with the spectral measures. In the lower panel, the average
flux-ratio of IUE/GALEX is shown as black diamonds connected by solid lines,
while the black and red filled circles connected by dotted lines are the results
from the two individual GALEX observations. The GALEX plus IUE statistical
uncertainties combined in quadrature appear as small black and red diamonds
connected by dashed lines with a scale in percent indicated on the right axis.
\label{cfiuehz4}} \end{figure}

An example for HZ4 of a star with two good GALEX data sets appears in
Figure~\ref{cfiuehz4}. In the NUV, the second GALEX observation (red) lies
below the first (black) shortward of 2600~\AA\ but shoots up abruptly longward
of 2600~\AA\ because of a slightly erroneous wavelength assignment before the
application of the steep GALEX sensitivity function. In other words, an
astrometric error makes the wavelengths a bit high in this case, which  causes
smaller sensitivities to be assigned with the largest impact in the region of
the sharp effective area decrease longward of 2600~\AA. HZ4 illustrates our
worst case for this occasional GALEX processing error, which affects our final
grand average correction by $\ll1\%$. The uncertainty of the abrupt upturn
(red diamonds in the lower panel) increases in correspondence with the
erroneous flux upturn and becomes larger than the uncertainty of the better
observation (black diamonds), so the weighted average GALEX correction (large
black diamonds) is very near the more reliable first observation (black dots).
The final IUE/GALEX flux ratio for HZ4 is an average which uses the combined
uncertainties of GALEX and IUE as weights. Because there are no valid GALEX
bins in the second observation (red circles) in the  FUV longward of 1525~\AA,
the average correction for HZ4 is defined only by the first (black) dataset
from 1575--1775~\AA.

Shortward of $\sim$2600~\AA\ in the regions of the best sensitivities,
the repeatability is $\sim$10\% between the separate GALEX observations (red
and black squares in Figure~\ref{cfiuehz4}). This Figure~\ref{cfiuehz4}, a
similar plot for Q1302-102, and the scatter among the Figure~\ref{iueavg} stars
suggest that GALEX spectrophotometry has a typical repeatability of 10--20\%.

\begin{figure}			
\centering 
\includegraphics*[height=7.5in]{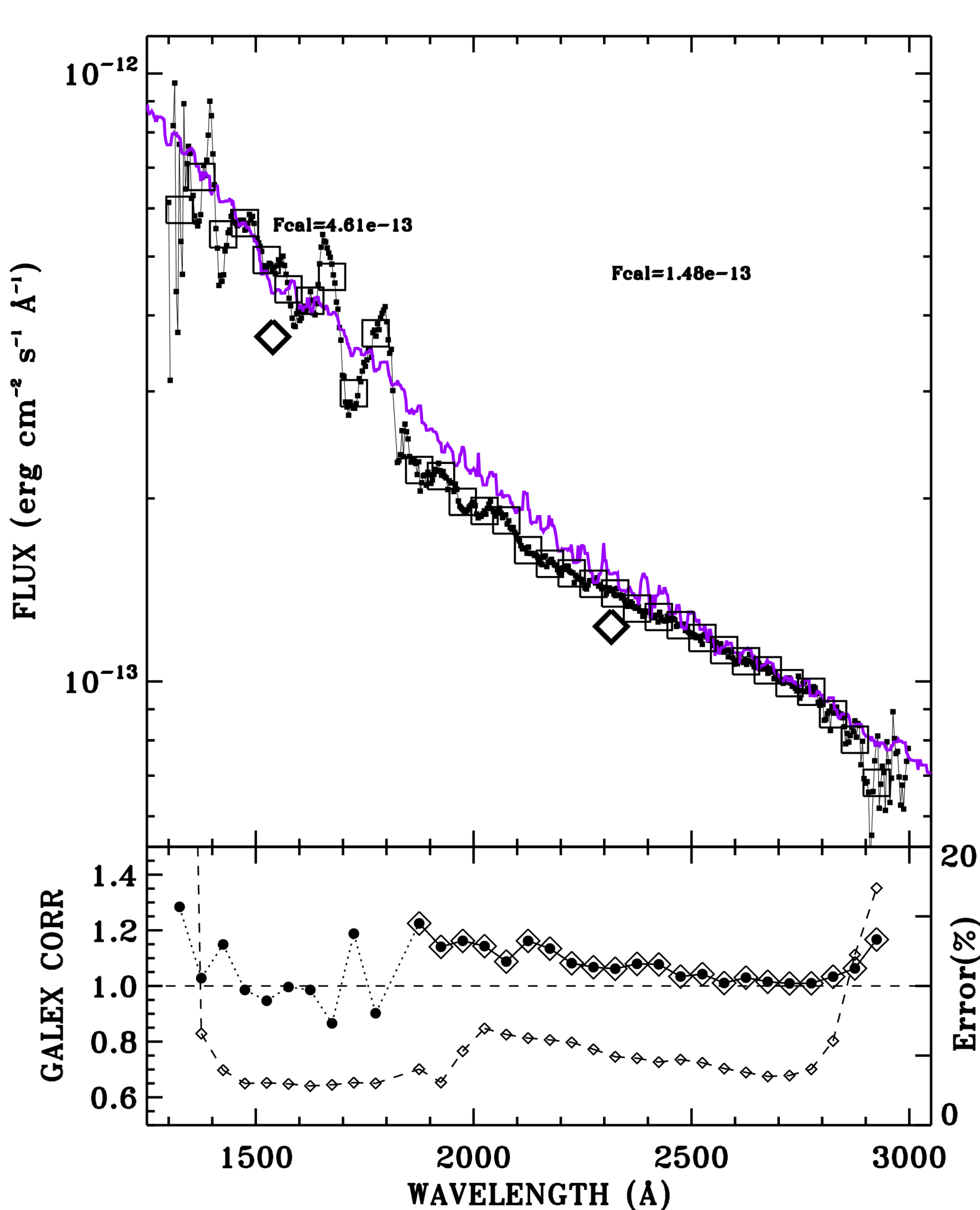}
\caption{\baselineskip=12pt
IUE spectrum of GD108 (purple) overplotted with GALEX spectra as small squares and
as large squares in 50~\AA\ bins, as in Figure~\ref{cfiuehz4}. The mean FUV IUE
flux at 1539~\AA\ is Fcal=4.61e$^{-13}$, which is slightly above the adopted
limit of 4.5e$^{-13}$ for good GALEX spectra.
In the lower panel, the ratio of IUE/GALEX fluxes is shown as black circles
and diamonds connected by solid lines, while the GALEX plus IUE statistical
uncertainties combined in quadrature appear as small black diamonds connected by
dashed lines with a scale in percent indicated on the right axis.
\label{cfiuegd108}} \end{figure}

An example of a star with flux just above our FUV limit of
4.5e$^{-13}$~erg s$^{-1}$ cm$^{-2}$ \AA$^{-1}$ appears in
Figure~\ref{cfiuegd108}. The choppy nature of the GALEX flux distribution
in regions of low sensitivity with a jump of $\sim$30\% from the
1675 to the 1725~\AA\ bin could be caused by unflagged overlapping spectra,
astrometric errors, or a cosmetic detector artifact. Consistent with
a variable saturation limit is GD50 with a 1539~\AA\ flux of 7.34e$^{-13}$ from
Table~\ref{table:galx}. The largest deviation from unity in the GD50 IUE/GALEX
FUV flux-ratio is $\sim$10\%, but the FUV GALEX data for GD50 is still excluded
for consistency.

~\\
~\\

\section{Conclusions} 		
\label{s_result}

\begin{figure}			
\centering 
\includegraphics*[height=7.5in]{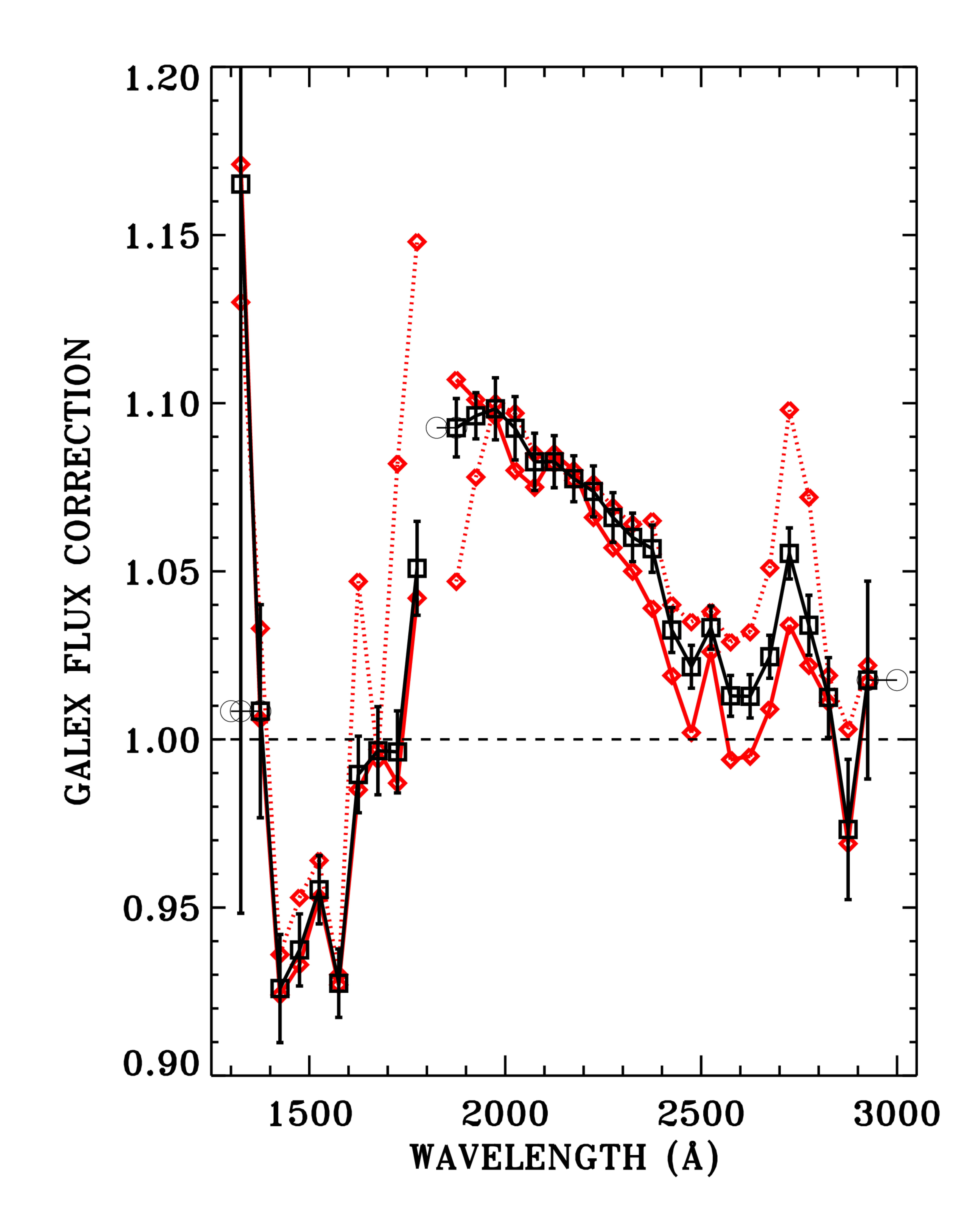}
\caption{\baselineskip=12pt
Average ratio of fluxes on the CALSPEC scale to GALEX fluxes. Red
diamonds connected by a dotted line are the results from a direct comparison of
CALSPEC to GALEX from Section 2, while the red diamonds connected by a solid
line are the ratios from Section 3 of the corrected IUE fluxes to GALEX. The
black squares with 1$\sigma$ error bars are the weighted average correction.
GALEX fluxes should be multiplied by this average correction from
Table~\ref{table:corr} to be on the HST/CALSPEC scale. Open circles near the
endpoints of the corrections are explained in the text.
\label{galxcor}} \end{figure}

For stars below our adopted saturation limits of 4.5e$^{-13}$~erg
s$^{-1}$ cm$^{-2}$ \AA$^{-1}$, Figure~\ref{galxcor} shows the correction
factors derived from the separate comparisons of GALEX with CALSPEC (red
diamonds connected by a dotted line) and IUE (red diamonds connected by a solid
line) and the final weighted average GALEX spectroscopic correction in
black. In the FUV, the CALSPEC/GALEX uncertainties are all larger than the
IUE/GALEX uncertainties, so the final results (black squares) are dominated by
the IUE correction. However in the NUV, the CALSPEC results tend to have the
lower uncertainty shortward of  2625~\AA. At the 1325~\AA\ bin, the error bar
is so large and the value (black square) is so extreme that the weighted
correction is replaced with the more conservative guess of assigning the
1375~\AA\ value of 1.008 at 1325~\AA. The endpoint bins of 1325 and 2925~\AA\
are extended to 1300 and 3000~\AA, respectively, to achieve full coverage of
the GALEX spectral range. Finally, the NUV average at 1875 is extended
shortward to 1825~\AA. These four extension points appear as open circles in
Figure~\ref{galxcor}.

Table~\ref{table:corr} contains the values of the multiplicative GALEX
spectroscopic flux correction, which can be interpolated to find
intermediate correction values. In the 1775--1825~\AA\ gap between the FUV and
NUV, interpolation is still adequate for the rare valid GALEX data point that
might fall in that gap. The formal statistical uncertainties in 
Table~\ref{table:corr} may be too optimistic, especially in the regions where
the independent CALSPEC and IUE corrections, i.e. red points in
Figure~\ref{galxcor}, disagree by more than the size of the black 1$\sigma$
error bars. For example, in the NUV longward of 2350~\AA, 3\% is a more
conservative estimate of the uncertainty of the correction. For any single GALEX
spectrum, the precision of the corrected fluxes is limited by the 10-20\%
repeatability floor for a single observational visit. Statistically, an average
of at least 10 GALEX visits is required before a 3\% uncertainty in the average
flux correction becomes important.

\section{Electronically Available Results} 		
\label{s_el}

The co-added and merged IUE spectra for the 18 stars of Table~\ref{table:galx}
and Table~\ref{table:iuestub} along with  the Table~\ref{table:corr} GALEX
correction factors are available in the on-line version of this paper, in MAST
as High-Level Science Products 
(HLSP)\footnote{\url{https://archive.stsci.edu/prepds/galex-fluxcal/}}, and in
the $uvsky$ project web site\footnote{\url{http::/dolomiti.jhu.edu/uvsky/}}. The
IUE merged ascii files of Table~\ref{table:iuestub} combine the datasets
described in Section \ref{s_iue} and cross from SWP to the long wavelength (LW)
cameras at 1975~\AA. These merged files contain eight columns: (1) wavelength in
\AA, (2) the average net signal in linearized IUE Flux Number (FN) units per
second, (3) the average flux in erg cm$^{-2}$ s$^{-1}$ \AA$^{-1}$, (4) the 
background signal in the same units as the net, (5) the formal propagated
uncertainty as the error-in-the-mean in flux units, (6) the number of
observations averaged, (7) the total exposure time in seconds, and (8) the rms
scatter among the observations in percent. Our next paper will make
electronically available the useful portion of the GALEX spectral collection of
over 100,000 SEDs corrected to the CALSPEC scale. More details of our work along
with the machine readable files also appear on the $uvsky$ project
website\footnote{\url{http://dolomiti.pha.jhu.edu/uvsky/}}.

Table~\ref{table:corr} and the IUE merged spectra from Table~\ref{table:iuestub} are available at \dataset[10.17909/T9DX2D {https://doi.org/10.17909/T9DX2D}. The GALEX GR6/7 data release on which this article is based is available at \dataset[10.17909/T9H59D]{https://doi.org/10.17909/T9H59D}.

\section{Summary}		
\label{s_sum}

Our comparison of GALEX UV-grism fluxes to the HST-based absolute flux standards
in CALSPEC and to corrected IUE SEDs \citep{b&b2018} produces a correction to
the GALEX spectroscopic flux scale. The correction is
wavelength-dependent and covers the GALEX range of sensitivity from 1300 to
3000~\AA. To be on the HST/CALSPEC flux scale, GALEX archival fluxes should be
multiplied by these Table~\ref{table:corr} correction factors that range from
0.926 to 1.098. With this flux correction, the fainter sources  that will be
available in our upcoming catalog of GALEX spectra will support 
cross-calibration and planning of observations with future UV space missions.

\section*{Acknowledgements}

Karen Levy, Bernie Shiao, and Randy Thomson provided many helpful suggestions
and clarifications with regard to the GALEX MAST database. Scott Fleming entered
our HLSP into MAST. Support for this work
was provided by NASA through the Space Telescope Science Institute, which is
operated by AURA, Inc., under NASA contract NAS5-26555. LB acknowledges support
from NASA grant NNX16AF40G. This research made use of the SIMBAD database,
operated at CDS, Strasbourg, France.

\textbf{ORCID iDs}

Ralph C. Bohlin https://orcid.org/0000-0001-9806-0551 and
Luciana Bianchi https://orcid.org/0000-0001-7746-5461

\newpage

\newpage

\newpage

\begin{deluxetable}{ccc}     
\tabletypesize{\scriptsize}
\tablewidth{0pt}
\tablecolumns{3}
\tablecaption{\label{table:corr} Multiplicative Correction for GALEX Fluxes}
\tablehead{
\colhead{Wavelength (\AA)} &\colhead{Correction} &\colhead{Uncertainty}}
\startdata
1300  &1.008  & ...    \\
1325  &1.008  & ...    \\
1375  &1.008  & 0.032 \\
1425  &0.926  & 0.016 \\
1475  &0.937  & 0.011 \\
1525  &0.955  & 0.010 \\
1575  &0.927  & 0.010 \\
1625  &0.990  & 0.011 \\
1675  &0.997  & 0.013 \\
1725  &0.996  & 0.012 \\
1775  &1.051  & 0.014 \\
1825  &1.093  & ...   \\
1875  &1.093  & 0.009 \\
1925  &1.096  & 0.007 \\
1975  &1.098  & 0.009 \\
2025  &1.093  & 0.009 \\
2075  &1.083  & 0.008 \\
2125  &1.083  & 0.008 \\
2175  &1.078  & 0.007 \\
2225  &1.074  & 0.008 \\
2275  &1.066  & 0.007 \\
2325  &1.060  & 0.007 \\
2375  &1.057  & 0.007 \\
2425  &1.033  & 0.007 \\
2475  &1.022  & 0.006 \\
2525  &1.033  & 0.006 \\
2575  &1.013  & 0.006 \\
2625  &1.013  & 0.006 \\
2675  &1.025  & 0.006 \\
2725  &1.055  & 0.008 \\
2775  &1.034  & 0.009 \\
2825  &1.012  & 0.012 \\
2875  &0.973  & 0.021 \\
2925  &1.018  & 0.029 \\
3000  &1.018  & ...   \\
\enddata
\tablecomments{Table~\ref{table:corr} is published in its entirety in 
machine-readable format.}
\end{deluxetable}

\begin{deluxetable}{cccccccc}     
\tablewidth{0pt}
\tablecolumns{8}
\tablecaption{\label{table:iuestub} IUE SEDs for 18 Stars}
\tablehead{
\colhead{Wavelength (\AA)} &\colhead{Net (s$^{-1}$)}
	&\colhead{Flux\tablenotemark{a}} &\colhead{Bkg (s$^{-1}$)}
	&\colhead{Sigma\tablenotemark{a}} &\colhead{No. Obs}
	&\colhead{Exp (s)} &\colhead{RMS (\%)} }
\startdata
BPM16274 \\
 1152.26  &2.649e-02  &2.564e-13  &5.917e-02  &1.928e-14  &5.0  &25197.9 &13.45\\
 1153.93  &2.589e-02  &2.353e-13  &5.150e-02  &1.562e-14  &7.0  &33596.9 &14.44\\
 1155.61  &3.330e-02  &2.827e-13  &5.153e-02  &1.568e-14  &7.0  &33596.9 &11.31\\
 1157.29  &2.781e-02  &2.198e-13  &5.156e-02  &1.373e-14  &7.0  &33596.9 &15.01\\
 1158.96  &3.432e-02  &2.507e-13  &5.159e-02  &1.334e-14  &7.0  &33596.9 &12.41\\
 1160.64  &3.837e-02  &2.584e-13  &5.162e-02  &1.270e-14  &7.0  &33596.9 & 9.57\\
\enddata
\tablenotetext{a}{erg s$^{-1}$ cm$^{-2}$ \AA$^{-1}$}
\tablecomments{Table~\ref{table:iuestub} is published in its entirety in a
machine-readable format which includes the SEDs for all 18 stars
of Table~\ref{table:galx}. A portion is shown here for guidance regarding
its form and content.}
\end{deluxetable}

\bibliographystyle{apj}

\bibliography{../../pub/paper-bibliog}

\begin{thebibliography}{}
\expandafter\ifx\csname natexlab\endcsname\relax\def\natexlab#1{#1}\fi

\bibitem[{{Barger} {et~al.}(2012){Barger}, {Cowie}, \& {Wold}}]{Barger2012}
{Barger}, A.~J., {Cowie}, L.~L., \& {Wold}, I.~G.~B. 2012, \apj, 749, 106

\bibitem[{{Bertone} \& {Chavez}(2011)}]{bertone2011}
{Bertone}, E., \& {Chavez}, M. 2011, \apss, 335, 69

\bibitem[{{Bianchi}(2012)}]{bianchi2012pn}
{Bianchi}, L. 2012, in IAU Symposium, Vol. 283, IAU Symposium, 45--52

\bibitem[{{Bianchi} {et~al.}(2018{\natexlab{a}}){Bianchi}, {de la Vega},
  {Shiao}, \& {Bohlin}}]{bianchi2018}
{Bianchi}, L., {de la Vega}, A., {Shiao}, B., \& {Bohlin}, R.
  2018{\natexlab{a}}, \apss, 363, 56

\bibitem[{{Bianchi} \& {Garcia}(2002)}]{bianchigarcia02}
{Bianchi}, L., \& {Garcia}, M. 2002, \apj, 581, 610

\bibitem[{{Bianchi} \& {Garcia}(2014)}]{bianchigarcia14}
---. 2014, Advances in Space Research, 53, 973

\bibitem[{{Bianchi} {et~al.}(2018{\natexlab{b}}){Bianchi}, {Keller}, {Bohlin},
  {Barstow}, \& {Casewell}}]{bianchi2018b}
{Bianchi}, L., {Keller}, G.~R., {Bohlin}, R., {Barstow}, M., \& {Casewell}, S.
  2018{\natexlab{b}}, \apss, 363, 166

\bibitem[{{Bianchi} {et~al.}(2012){Bianchi}, {Manchado}, \&
  {Forster}}]{bianchi2012pn2}
{Bianchi}, L., {Manchado}, A., \& {Forster}, K. 2012, in IAU Symposium, Vol.
  283, IAU Symposium, 308--309

\bibitem[{{Bohlin} \& {Bianchi}(2018)}]{b&b2018}
{Bohlin}, R.~C., \& {Bianchi}, L. 2018, \aj, 155, 162

\bibitem[{{Bohlin} {et~al.}(2014){Bohlin}, {Gordon}, \&
  {Tremblay}}]{bohlinetal14}
{Bohlin}, R.~C., {Gordon}, K.~D., \& {Tremblay}, P.-E. 2014, \pasp, 126, 711
  (B14)

\bibitem[{{Bohlin} \& {Koester}(2008)}]{bohlin-k2008}
{Bohlin}, R.~C., \& {Koester}, D. 2008, \aj, 135, 1092

\bibitem[{{Burgarella} {et~al.}(2005){Burgarella}, {Buat}, {Small}, {Barlow},
  {Boissier}, {Gil de Paz}, {Heckman}, {Madore}, {Martin}, {Rich}, {Bianchi},
  {Byun}, {Donas}, {Forster}, {Friedman}, {Jelinsky}, {Lee}, {Malina},
  {Milliard}, {Morrissey}, {Neff}, {Schiminovich}, {Siegmund}, {Szalay},
  {Welsh}, \& {Wyder}}]{Burgarella2005}
{Burgarella}, D., {Buat}, V., {Small}, T., {et~al.} 2005, \apjl, 619, L63

\bibitem[{{Cowie} {et~al.}(2010){Cowie}, {Barger}, \& {Hu}}]{Cowie2010}
{Cowie}, L.~L., {Barger}, A.~J., \& {Hu}, E.~M. 2010, \apj, 711, 928

\bibitem[{{Deharveng} {et~al.}(2008){Deharveng}, {Small}, {Barlow},
  {P{\'e}roux}, {Milliard}, {Friedman}, {Martin}, {Morrissey}, {Schiminovich},
  {Forster}, {Seibert}, {Wyder}, {Bianchi}, {Donas}, {Heckman}, {Lee},
  {Madore}, {Neff}, {Rich}, {Szalay}, {Welsh}, \& {Yi}}]{Deharveng2008}
{Deharveng}, J.-M., {Small}, T., {Barlow}, T.~A., {et~al.} 2008, \apj, 680,
  1072

\bibitem[{{Gal-Yam} {et~al.}(2008){Gal-Yam}, {Bufano}, {Barlow}, {Baron},
  {Benetti}, {Cappellaro}, {Challis}, {Ellis}, {Filippenko}, {Foley}, {Fox},
  {Hicken}, {Kirshner}, {Leonard}, {Li}, {Maoz}, {Matheson}, {Mazzali},
  {Modjaz}, {Nomoto}, {Ofek}, {Simon}, {Small}, {Smith}, {Turatto}, {Van Dyk},
  \& {Zampieri}}]{Gal-Yam2008}
{Gal-Yam}, A., {Bufano}, F., {Barlow}, T.~A., {et~al.} 2008, \apjl, 685, L117

\bibitem[{{Garcia} \& {Bianchi}(2004)}]{garciabianchi04}
{Garcia}, M., \& {Bianchi}, L. 2004, \apj, 606, 497

\bibitem[{{Godon} {et~al.}(2014){Godon}, {Sion}, {Starrfield}, {Livio},
  {Williams}, {Woodward}, {Kuin}, \& {Page}}]{godon2014}
{Godon}, P., {Sion}, E.~M., {Starrfield}, S., {et~al.} 2014, \apjl, 784, L33

\bibitem[{{Herald} \& {Bianchi}(2007)}]{heraldbianchi07}
{Herald}, J.~E., \& {Bianchi}, L. 2007, \apj, 661, 845

\bibitem[{{Herald} \& {Bianchi}(2011)}]{heraldbianchi11}
---. 2011, \mnras, 417, 2440

\bibitem[{{Joyce} {et~al.}(2018){Joyce}, {Barstow}, {Casewell}, {Burleigh},
  {Holberg}, \& {Bond}}]{joyce2018}
{Joyce}, S.~R.~G., {Barstow}, M.~A., {Casewell}, S.~L., {et~al.} 2018, \mnras,
  479, 1612

\bibitem[{{Montez} {et~al.}(2017){Montez}, {Ramstedt}, {Kastner}, {Vlemmings},
  \& {Sanchez}}]{montez2017}
{Montez}, Jr., R., {Ramstedt}, S., {Kastner}, J.~H., {Vlemmings}, W., \&
  {Sanchez}, E. 2017, \apj, 841, 33

\bibitem[{{Morrissey} {et~al.}(2007){Morrissey}, {Conrow}, {Barlow}, {Small},
  {Seibert}, {Wyder}, {Budav{\'a}ri}, {Arnouts}, {Friedman}, {Forster},
  {Martin}, {Neff}, {Schiminovich}, {Bianchi}, {Donas}, {Heckman}, {Lee},
  {Madore}, {Milliard}, {Rich}, {Szalay}, {Welsh}, \& {Yi}}]{morrissey2007}
{Morrissey}, P., {Conrow}, T., {Barlow}, T.~A., {et~al.} 2007, \apjs, 173, 682

\bibitem[{{Nichols} \& {Linsky}(1996)}]{nichols1996}
{Nichols}, J.~S., \& {Linsky}, J.~L. 1996, \aj, 111, 517

\bibitem[{{Pala} {et~al.}(2015){Pala}, {G{\"a}nsicke}, {Beuermann}, {Bildsten},
  {De Martino}, {Godon}, {Henden}, {Hubeny}, {Knigge}, {Long}, {Marsh},
  {Patterson}, {Schreiber}, {Sion}, {Szkody}, {Townsley}, \&
  {Zorotovic}}]{pala2015}
{Pala}, A.~F., {G{\"a}nsicke}, B.~T., {Beuermann}, K., {et~al.} 2015, in
  Astronomical Society of the Pacific Conference Series, Vol. 493, 19th
  European Workshop on White Dwarfs, ed. P.~{Dufour}, P.~{Bergeron}, \&
  G.~{Fontaine}, 521

\bibitem[{{Valencic} {et~al.}(2004){Valencic}, {Clayton}, \&
  {Gordon}}]{valencic2004}
{Valencic}, L.~A., {Clayton}, G.~C., \& {Gordon}, K.~D. 2004, \apj, 616, 912

\bibitem[{{Wold} {et~al.}(2017){Wold}, {Finkelstein}, {Barger}, {Cowie}, \&
  {Rosenwasser}}]{Wold2017}
{Wold}, I.~G.~B., {Finkelstein}, S.~L., {Barger}, A.~J., {Cowie}, L.~L., \&
  {Rosenwasser}, B. 2017, \apj, 848, 108

\end{thebibliography}

\end{document}